\documentclass[reprint,amsmath,amssymb,aps,pra]{revtex4-1}

\begin{document}

\title{Born Rule and   Noncontextual Probability}

\author{Fabrizio Logiurato$^1$ and Augusto Smerzi$^{1,2}$}
\affiliation{
$^1$INO-CNR BEC Center and Physics Department, Trento University, I-38123 Povo, Italy \\
$^2$INO-CNR and LENS, 50125 Firenze, Italy}

\date{19 February 2012}

\begin{abstract}
\noindent
The probabilistic rule that links the formalism of Quantum Mechanics (QM) to the real world was stated by Born in 1926. 
Since then, there were many attempts to derive the Born postulate as a theorem,
Gleason's  being the most prominent. The Gleason derivation, however, is generally 
considered  rather 
intricate and its physical meaning, in particular in relation with the noncontextuality of probability (NP),
is not quite evident. 
More recently, we are witnessing a revival of interest in possible demonstrations of the Born rule,
like Zurek's and Deutsch's based on the decoherence and on the theory of decisions, respectively. 
Despite an ongoing debate about the presence of hidden assumptions and 
circular reasonings, these have the merit of 
prompting more physically oriented approaches to the problem.

Here we suggest a new proof of the Born rule based on the noncontextuality of probability.
Within the theorem we also demonstrate the continuity of probability with respect to the amplitudes,
which has been suggested to be a gap in Zurek's and Deutsch's approaches, and we show
that NP is
implicitly postulated also in their demonstrations. Finally, physical motivations of NP  are given  based on an invariance principle with respect to a resolution change  of measurements and with respect to the principle of no-faster-than-light signalling. 

\end{abstract}\pacs{}
\maketitle


\section{Introduction}
The fundamental probabilistic postulate which allows QM its successful predictions was first introduced by Born (from a suggestion by Einstein) in 1926 within the scattering theory \cite{Born1926}.  
The question if such postulate could be derived as a theorem from  other postulates of QM 
was quickly posed. A first answer was provided by 
Von Neumann and included in his famous book of 1932 on the mathematical foundations of QM \cite{vonNeumann}. His theorem nowadays does not enjoy much attention since it is believed to
contain more assumptions than Gleason's \cite{Jordan}. Different attempts
have been discussed in the context of the relative-state \cite{Everett1957} and the many-worlds interpretations \cite{DeWitt1971} of QM by
Finkelstein \cite{Finkelstein1965} and Hartle \cite{Hartle1968}, which have
analyzed an endless sequence of measurements to show that the relative frequency follows the Born rule. However the meaning in the real world of
an infinite sequence of measurements is controversial
and if these proofs contain circularities is not clear \cite{Graham1973}-\cite{Kent1990}. 

The 1957 Gleason work \cite{Gleason1957} is regarded as the most important demonstration of the Born rule \cite{Caves2002}. Yet, this theorem is usually considered  quite formal and difficult to grasp. Moreover, it is based on the concept of noncontextuality of probability (NP), whose connection with physics is not clear. 
In the structure of the Gleason theorem, every Hilbert subspace $\mathcal M$  corresponds to an observable quantity. Each one of such  subspaces is representable by a projection operator $\hat{\Pi}_{\mathcal M}$ and every  orthonormal basis of vectors $\{|a_{i}\rangle \}$  is related to a complete  mutually set of commuting projectors $\{ \hat{\Pi}_i=|a_{i}\rangle \langle a_{i} | \}$ .  
A projector $ \hat{\Pi}=|\psi\rangle \langle \psi | $ can represent the question with the answer yes-no  when we make an experiment testing if the system has the respective state $|\psi\rangle$. A set of commuting projectors corresponds to a set of questions which can simultaneously be asked in a measurement.
In the logical-algebraic approach \cite{Redhead}, probability is a measure defined over a projection lattice (set of closed subspaces) of Hilbert space, a mapping 
$p:\hat{\Pi}_{\mathcal M} {\rightarrow}[0,1]$. The postulates on which the theorem rests are:
\begin{enumerate}
\item The probability assigned to a complete set of projectors is normalized, if $dim \mathcal H=N$:
\begin{equation} \label{npsg}
\sum_{i=1}^{N} p(\hat{\Pi}_i ) =1 \,,  \qquad   p(\hat{\Pi}_i )\in [0,1]    \,.
\end{equation}
\item For any sequence of mutually orthogonal projectors:  
\begin{equation} \label{npsgnp}
p( \sum_{i=1}^{M}\hat{\Pi}_i )=\sum_{i=1}^{M} p(\hat{\Pi}_i )  \,, \qquad    M\leq N       \,.
\end{equation}
\end{enumerate}
\noindent
Postulate 2 is equivalent to assuming NP: 
the probability of  a certain occurrence, for example the answer yes for a projector 
does not depend on  other questions simultaneously tested, that is, on  other projectors. In terms of basis vectors, 
the probability of obtaining a given state  is independent of the basis it belongs to. 
To illustrate this important point, let us consider, 
for instance, two complete set of mutually orthogonal projectors $\{\hat{\Pi}_i\}$, $\{\hat{\Pi}'_i\}$  of a Hilbert space  with $dim \mathcal H=3$, where $\hat{\Pi}_1 \equiv\hat{\Pi}'_1$.
From the normalization postulate of probability:
\begin{subequations} \label{pnpp}
\begin{align}
p(\hat{\Pi}_1)+p(\hat{\Pi}_2)+p(\hat{\Pi}_3)   &=1   \label{pnppa}      \,,      \\[2.5ex]
p'(\hat{\Pi}_1)+p(\hat{\Pi}'_2)+p(\hat{\Pi}'_3)  &=1   \label{pnppb}      \,,       
\end{align}
\end{subequations}
\noindent
where in principle $p(\hat{\Pi}_1)$  and $p'(\hat{\Pi}_1)$  may be different.
In the subspace orthogonal to $\hat{\Pi}_1$  we have $\hat{\Pi}_2+\hat{\Pi}_3=\hat{\Pi'}_2+\hat{\Pi'}_3$, then from  Postulate 2:
\begin{equation} \label{pbdp2}
p(\hat{\Pi}_2)+p(\hat{\Pi}_3)=p(\hat{\Pi}'_2)+p(\hat{\Pi}'_3)  \,, \qquad         
\end{equation}
\noindent
and by comparing  Eq. (\ref{pnpp}) with  Eq. (\ref{pbdp2}) we get  $p(\hat{\Pi}_1)=p'(\hat{\Pi}_1)$. It is also possible to show the vice versa, i.e. if NP holds then Postulate 2 is true.
Thus Postulate 2 is equivalent to postulate NP.
From the Postulates 1 and 2 it follows
\vskip.2cm
\noindent
{\sl Gleason Theorem}:  For $dim H \geq 3$,  there exists  an operator density $\hat{\rho}$ such that
the  probability measure for every projector $\hat{\Pi}$ is the rule trace:
\begin{equation} \label{trru}
p(\hat{\Pi})= tr (\hat{\rho}\hat{\Pi} )      \,,
\end{equation}
\noindent
namely, the Born rule.
\vskip.2cm

More recently, new proofs appeared in the literature which are 
``more physically motivated than Gleason's argument''  to demonstrate the Born rule.
Particularly interesting are Deutsch's and Zurek's derivations \cite{Deutsch1999}-\cite{Zurek2005}, also
for their further motivation to demonstrate 
the inherently probabilistic nature of QM and how
the meaning of probability emerges from the theory.
Such an ambitious goal, however, has been disputed 
and some authors suggest the presence of  circularity flaws.
According to \cite{Caves2005}-\cite{Barnum2003}, it is not possible
to derive the Born rule without introducing from the very beginning 
some assumptions about the existence of probability into the theory. 

Deutsch's and Zurek's demonstrations of the Born rule have similar structures. As a first step,
they use an invariance principle in order to show that equal amplitudes correspond to equal probabilities. 
Deutsch finds such principle in the theory of decisions 
(for critical discussions and further elaborations see \cite{Barnum2000}-\cite{Wallace2003}).
Zurek introduces an environment-assisted invariance principle \cite{Zurek2002b, Zurek2003c},
{\it envariance}, in the framework of a relative-state approach where the system $\mathcal S$ under measurement
jointly evolves with the environment
$\mathcal E$ and, possibly, with one or more auxiliary measurement devices. 
After demonstrating the correspondence between equal amplitudes and equal probabilities, 
both Zurek and Deutsch consider a fine-graining
technique to deal with the general case of different amplitudes of the initial state. 
For this they have to introduce auxiliary
systems which become entangled with  $\mathcal S$.

However fine-graining has been questioned: 
Caves \cite{Caves2005} considers disturbing the necessity of introducing
additional systems
of adequate dimensions, possibly in an infinite Hilbert space, in order to reach the wanted approximation to  irrational probabilities.
Barnum \cite{Barnum2003} points out an even more stringent weakness: the step from rational 
to irrational amplitudes requires
the continuity of the probability with respect to such amplitudes and such a property is not demonstrated by Zurek and Deutsch. 
We note that a fundamental and extensive part of Gleason's theorem 
is actually devoted to prove that the probability is continuous. 

In the next section we intend to demonstrate the Born rule from the NP assumption. 
In our approach we do not enter into the controversy about the origin and the meaning of probability,
and from the very beginning, following Gleason (see also
Mohrhoff in Ref. \cite{Mohrhoff}), we  assume the existence of probability 
as a measure definite on a quantum state.
In addition, we do not suppose unitary evolution and consider a single system,
therefore our theorem is very close to Gleason's original formulation. However, differently from Gleason, we consider as a postulate the  state of a system described by a vector in the Hilbert space, which allows a
much more elementary deduction of the Born rule. As  Gleason's theorem, our result holds for a
dimension of the Hilbert space $N\geq 3$, where $N$ could be finite. 
We also give a rather simple demonstration that the probability is continuous from  NP. Without this proof,  which  takes up a conspicuous part of Gleason's work, we could state that Deutch's and Zurek's 
results as well as ours, hold only for rational values of amplitudes.

In Section III we generalize our theorem to include entangled states 
and multi-particles systems. Since the Hilbert space $\mathcal H$ of such systems has dimension 
which is the product of the dimensions of its components, the coupling of a subsystem 
with $dim {\mathcal H} \geq 3$ with another subsystem with 
$dim {\mathcal H} =2$ enables us to extend the Born rule also to such subspaces.
 
In  Section IV we show that the Gleason postulate of NP is identifiable in the use of fine-graining
and that such a postulate is also a part of Zurek's and Deutsch's proofs. Indeed,
fine-graining is a resolution change of our measurements and 
 NP is equivalent to a condition of invariance 
with respect to such a change: fine-graining 
on a subspace of the state must not alter the probability of finding the system in a different subspace.
This resolution invariance is an intuitive and physically natural interpretation of NP.

Finally, in  Section V, we  show that
the condition of noncontextual probability, or resolution invariance, can be also
deduced by a hypothesis of no-faster-than-light signalling.


\section{Born rule as a theorem from noncontextual probability}
Contrarily to the principles of classical mechanics, the axioms of QM do not have an equally shared and clear formulation. 
We do not want to enter into such a thorny issue, related to the interpretative problems of QM. For our purpose, 
from the usually regarded quantum postulates in standard presentations of QM (see, e.g., \cite{CohenT} - \cite{Galindo}), we can extract the 
following common assumptions:
\vskip.3cm
\noindent
{\sl Postulate I}: At the time $t$  the state of every physical system $\mathcal S$ is described by a normalized vector
$|\psi_S (t) \rangle$  belonging to a Hilbert space $\mathcal{H}$. 
\vskip.3cm
\noindent
{\sl Postulate II}: Every  measurable  physical quantity $\mathcal A$  is described by a  Hermitian operator $\hat{A}$ acting in $\mathcal{H}$.
\vskip.3cm
\noindent
{\sl Postulate III}: The only possible result of a measurement  of an observable $\mathcal A$  is one of  the eigenvalues  $\left\{a_i \right\}$ of 
$\hat{A}$, (as the customary habit, from now on we will identify the observable with its operator).
\vskip.5cm
\noindent
{\sl Postulate I} is  known as the {\sl Completeness Postulate of QM}. {\sl Postulate II} links an observable quantity with an operator in a Hilbert space  
({\sl observable-operator link}).
{\sl Postulate III} connects a particular value of that observable with an eigenvalue of the corresponding operator {(\sl value-eigenvalue link}).
\vskip.3cm
Now consider  an observable $\hat{A}$ with discrete eigenvalues $\{a_i\}$ and eigenstates
$\{|a_i \rangle \}$, $i=1, \ldots   ,N$,  and a single system $\mathcal S$ in the state $|\psi_S \rangle \in \mathcal{H}^N$. $\{|a_i \rangle \}$ is a complete orthonormal set and $|\psi_S \rangle$ can be decomposed into components:
\begin{equation} \label{ps}
|\psi_S \rangle= \sum_{i=1}^{N} c_{i}| a_i \rangle  \,.
\end{equation}
\noindent
We intend to prove the following
\vskip.3cm
\noindent
{\sl Theorem 1}: If NP and the postulates of QM hold (except for the Born rule and the unitary dynamics), then for $N\geq3$ the probability  of obtaining the non-degenerate eigenvalue  $a_i$,  measuring $\hat{A}$ on the state $|\psi_S \rangle$, is necessarily given by  the square modulus of the inner product:
\begin{equation} \label{pib}
p(a_i)=\left| \langle a_i | \psi_S (t) \rangle \right|^2   \quad \text{ Born Rule} \,,
\end{equation}
\noindent
where $| a_i \rangle$ is the eigenstate associated with the eigenvalue $a_i$.
\vskip.4cm
\noindent
We assume the probability as a primitive concept and we begin the demonstration of {\sl Theorem 1} by
proving some simple lemmas. 
We underline that any theorem having the aim to prove the Born rule  has to
face the fact that the connection between probability and quantum state 
must be postulated or deduced by other 
means. Gleason, for instance, postulates  the probability to be a measure definite on
states of the basis. 

As Gleason, we suppose QM to be a probabilistic theory, which makes predictions about probabilities of
occurrences of observable quantities, and that  probabilities are functions of  quantum states. 
Then, from QM postulates we want to show that
\vskip.4cm
\noindent
{\sl Lemma 1}: The probability of obtaining an eigenvalue $a_i$ is equal to the
probability of obtaining the respective eigenstate $| a_i \rangle$ (statistical case):
\begin{equation} \label{pia}
p(a_i)= p(| a_i \rangle)       \,.
\end{equation}
\vskip.2cm
\noindent
{\sl Proof}: Suppose the system is prepared with the eigenvalue $a_i$. In this case $p(a_i)=1$. 
According to the postulate of QM ({\sl eigenvalue-eigenstate link}):
\vskip.3cm
\noindent
{\sl Postulate IV}: If  the  measurement  of  the observable  $\hat{A}$ on the  system in the state $|\psi_S (t) \rangle$  gives the result  $a_i$, the state of the system immediately after the measurement  is the associated normalized eigenstate  $| a_i \rangle$ (non-degenerate case). 
\vskip.3cm
\noindent
Hence if $p(a_i)=1$ then $p(|a_i\rangle)=1$, and vice versa.
But $\left\{a_i\right\}$ correspond to mutually exclusive events,
thus has to be $p(a_j)=0$ for $j \neq i$. Then must necessarily be  $p(|a_j\rangle)=0$ too, since,  otherwise,
having obtained with a measurement the state $|a_j\rangle$ we should assign to the system the eigenvalue $a_j$,
in contradiction with the exclusivity property of $\{a_i\}$ events. Therefore we have
the equation between  conditional probabilities (non-statistical case):
\begin{equation} \label{picond}
p(|a_i \rangle ; a_j)=p(a_i ; |a_j \rangle)= \delta_{ij}    \,.
\end{equation}
\noindent
Because of {\sl Postulate III} of QM, the only possible result of a measurement is one of the eigenvalues of $\hat{A}$. Then from the Bayesian formula of the joint probability, for a generic preparation $|\psi_S \rangle$ of the initial state, we have
\begin{equation} \label{pbeves}
\begin{array}{rl}
p(|a_i \rangle) &= {\displaystyle \sum_{j=1}^N} p(|a_i \rangle, a_j) \\
&={\displaystyle \sum_{j=1}^N} p(| a_i \rangle ;  a_j) p(a_j)
=p(a_i)            \,,
\end{array}
\end{equation}
\noindent
where we have used Eq. (\ref{picond}) in the conditional probabilities of Eq. (\ref{pbeves}). $\Box$
\vskip.4cm

We note from Eqs. (\ref{picond},\ref{pbeves}) that 
\begin{equation} \label{picondkk}
p(|a_i \rangle ; |a_j \rangle)= \delta_{ij}  \,,   \qquad   \sum_{i=1}^{N} p(| a_i \rangle)=1   \,.
\end{equation}

\noindent
{\sl Lemma 1} can be easily generalized for entangled states as well.
For instance, consider the  state:

\begin{equation} \label{psc}
|\psi_{SB} \rangle= \sum_{i=1}^{N} \sum_{k=1}^{M} c_{ik}| a_i \rangle | b_k\rangle \,,
\end{equation}

\noindent
where $\hat{B}$ is a new observable commuting with $\hat{A}$, with eigenvalues $\left\{b_k \right\}$ and eigenstates
$\left\{|b_k \rangle \right\}$,
$k=1, \ldots   ,M$,  and
$|\psi_{SB} \rangle  \in \mathcal{H}^N \otimes \mathcal{H}^M$, (we point out that
$\hat{B}$  does not  necessarily have to belong to an other system: for example, $\hat{A}$ could be the
spin observable and $\hat{B}$ the position observable  of one particle).
Consider the pairs of eigenvalues $\left\{a_i, b_k\right\}$,
and  eigenstates $\left\{|a_i \rangle , |b_k \rangle \right\}$. 
With the  trick of the conditional probabilities, it
follows that on the generic state $|\psi_{SB} \rangle$:

\begin{equation} \label{bfmac}
\begin{array}{rl}
p( | a_i \rangle  | b_k \rangle )&= p( |a_i \rangle , | b_k \rangle )   \\[2.5ex]
&=  p(\left\{a_i, b_k \right\})   \,,
\end{array}
\end{equation}

\noindent
(see appendix \ref{lemma1} for a proof of this).
Then the probability $p(a_i)$ is again given by:

\begin{equation} \label{pacb}
\begin{array}{rl}
p(a_i)&= {\displaystyle \sum_{k=1}^M} p(\left\{a_i, b_k \right\}) =
{\displaystyle \sum_{k=1}^M} p( | a_i \rangle | b_k \rangle )   \\[3ex]
& = {\displaystyle \sum_{k=1}^M} p( \{| a_i \rangle  , | b_k \rangle \}) =p( | a_i \rangle ) \,.
\end{array}
\end{equation}

\noindent
With a similar reasoning we can also show that $p(b_k)= p( | b_k \rangle )$.
We notice from  Eqs. (\ref{bfmac},\ref{pacb}) that detections  of
$\left\{ | a_i \rangle | b_k \rangle \right\}$   are mutually exclusive events as well,
with the condition of normalization which follows from the normalization of $p(a_i)$.

\vskip.4cm


\noindent
{\sl Lemma 2}: If a coefficient of the development in Eq. (\ref{ps}) is zero, the  probability of obtaining
the associated eigenvalue and eigenstate is zero as well.

\vskip.4cm

\noindent
{\sl Proof}:
If, for instance, $c_{k}=0$, the corresponding state $|a_k \rangle$ will be orthogonal to $|\psi_S \rangle$.
We can think $\{|\psi_S \rangle, |a_k \rangle \}$ as eigenstates of  the projector
 $\hat{\Pi}_{\mathcal S}=|\psi_S \rangle \langle \psi_S  |$ with eigenvalues $\left\{1,0\right\}$.
The Hermitian operator $\hat{\Pi}_{\mathcal S}$ represents the observable testing whether the system $\mathcal S$ is prepared with the state 
$|\psi_S \rangle$ or not. The pair $\{|\psi_S \rangle, |a_k \rangle \}$ is a complete orthonormal basis of a subspace 
$\mathcal H^{2}$ of $\mathcal H^{N}$.

With the system  prepared in the state $|\psi_S \rangle$: $p( |\psi_S \rangle ; |\psi_S \rangle)= 1$.
From  {\sl Lemma 1},  it follows that $p( 1 ; |\psi_S \rangle)= 1$ and because eigenvalues are exclusive
events $p( 0 ; |\psi_S \rangle)= 0$. So, if we apply {\sl Lemma 1} again to the previous equation,
considering that $|a_k \rangle$ is eigenstate of both  $\hat{\Pi}_{\mathcal S}$ and $\hat{A}$, we have $p(|a_k \rangle ; |\psi_S \rangle)= 0 $  
and $p(a_k ; |\psi_S \rangle)= 0$. $\Box$

\vskip.4cm

\noindent
Also {\sl Lemma 2 } can be directly generalized to entangled states. For instance, if in the state $|\psi_{SB} \rangle$
the coefficient $c_{kl}=0$ then   $p(|a_k \rangle |b_l \rangle   ; |\psi_{SB} \rangle)= 0$.
In particular, if the state of the system is in a Schmidt decomposition:

\begin{equation} \label{pscsd}
|\psi_{SB} \rangle= \sum_{i=1}^{N}  c_{i}| a_i \rangle | b_i\rangle \,,
\end{equation}

\noindent
from {\sl Lemma 1} and {\sl Lemma 2}, since
$p(| a_i \rangle | b_k\rangle)=0$ for $k\neq i$, follows the perfect correlation property:

\begin{equation} \label{pacbsd}
\begin{array}{rl}
p(a_i)= {\displaystyle \sum_{k=1}^M} p(\left\{a_i, b_k \right\}) &=
p( | a_i \rangle | b_i \rangle )= p( | a_i \rangle )   \\[3ex]
&=p(| b_i \rangle)=p(b_i)     \,.
\end{array}
\end{equation}


\vskip.4cm

\noindent
{\sl Lemma 3}: If the  probability is noncontextual the probability of obtaining the eigenvalue $a_i$ can depend only on the modulus
of the coefficient $c_i$:

\begin{equation} \label{pci}
p(a_i) =p(|c_i|) \,.
\end{equation}

\vskip.4cm

\noindent
{\sl Proof} : 
Let us recall the meaning of NP with an example.
Consider two operators  $\hat{A}$ and $\hat{A}'$ which
do not commute, $\left[\hat{A}, \hat{A}'\right] \neq 0$.
We take a Hilbert space with $dim \mathcal H=3$. Eigenvalues and eigenstates of $\hat{A}$ and $\hat{A}'$ are,
respectively:

\begin{equation} \label{Aee}
\begin{array}{rl}
\hat{A}  : & \left\{a_1, a_2, a_3 \right\} ,  \quad  \left\{ | a_1 \rangle, | a_2 \rangle, | a_3 \rangle\right\} , \\[2.5ex]
\hat{A}' : & \left\{a_1, a'_2, a'_3 \right\} ,  \quad  \left\{ | a_1 \rangle, | a'_2 \rangle, | a'_3 \rangle\right\} ,      \end{array}
\end{equation}

\noindent
where they have in common the eigenvalue $a_1$ and the associated eigenstate (we see that this is possible, if $\hat{A}$ and $\hat{A}'$ must
not commute,  just for $N\geq3$). We can write the state $|\psi_S \rangle $ in the two different bases of Eq. (\ref{Aee}):

\begin{equation} \label{psic}
\begin{array}{rl}
|\psi_S  \rangle &=   c_{1}|a_1 \rangle + c_{2}|a_2 \rangle + c_{3}|a_3 \rangle \\[2.5ex]
&=c_{1}|a_1 \rangle + c'_{2}|a'_2 \rangle + c'_{3}|a'_3 \rangle \,.
\end{array}
\end{equation}

\noindent
In general, the probability of obtaining the eigenvalue $a_1$ could depend on whether we measure  the observable $\hat{A}$ or $\hat{A}'$.
Probability could be different if we are in the experimental context  which is prepared for measuring which value of the triplet 
$\left\{a_1, a_2, a_3 \right\}$  $\hat{A}$ will have, or which value of the triplet
$\left\{a_1, a'_2, a'_3 \right\}$ $\hat{A}'$ will have.

With the result of {\sl Lemma 1} and the completeness postulate  of QM, we can write the more general
probability in different experimental contexts as:

\begin{equation} \label{pc}
\begin{array}{rl}
p_A(a_1) &= p(|a_1 \rangle ;|\psi_S  \rangle , \left\{ | a_1 \rangle, | a_2 \rangle, | a_3 \rangle\right\}) \,, \\[2.5ex]
p_{A'}(a_1) &= p(|a_1 \rangle ;|\psi_S  \rangle , \left\{ | a_1 \rangle, | a'_2 \rangle, | a'_3 \rangle\right\}) \,.
\end{array}
\end{equation}

\noindent
The probability is noncontextual if $p(a_1)$ does not depend on the measurement with other eigenvalues:

\begin{equation} \label{ncp}
p_{A}(a_1) \equiv p_{A'}(a_1) \,,
\end{equation}

\noindent
or, according to {\sl Lemma 1}, the probability $p(|a_1 \rangle)$ does not depend on the measurement with other eigenstates:

\begin{equation} \label{ncps}
p_{A}(|a_1 \rangle) \equiv p_{A'}(|a_1 \rangle) \,.
\end{equation}

\noindent
Then, by assuming NP, Eqs. (\ref{pc}) are reduced to the equation: 

\begin{equation} \label{ncp2}
p(a_1) = p(|a_1 \rangle ;|\psi_S \rangle)\,,
\end{equation}

\noindent
stating that the probability  only depends  on the eigenstate $|a_1 \rangle$ and, from the completeness postulate of QM
and NP, on the state of the system $|\psi_S \rangle$.

Thus, let us choose a new basis $\{|a'_i \rangle \}$ such that the state in  Eq. (\ref{ps}) is written with only  two components:

\begin{equation} \label{psit21}
|\psi_{S} \rangle =   c_{1}|a_1 \rangle  + |c'_{2}||a'_2\rangle       \,,
\end{equation}

\noindent
where $c'_i=0$ for $i\geq3$, $|c'_{2}|$ is the modulus of the amplitude $c'_{2}$ and

\begin{equation}
 \label{btnc}
|a'_1 \rangle \equiv|a_1 \rangle  \,, 
\qquad |a'_2 \rangle =\frac{1}{|c'_2|} \sum_{i=2}^N  c_i|a_i \rangle   \,.
\end{equation}

\noindent
$\{|a_1 \rangle ,|a'_2 \rangle \}$ could be regarded as  eigenstates of an observable $\hat{A}'=|a_1\rangle \langle a_1 |$ with eigenvalues 
$\{ 1,0 \}$ checking if our system is   in the state $|a_1 \rangle$ or not.

According to NP the transformation of basis $\{|a_i \rangle \} \rightarrow \{|a'_i \rangle \}$ of the state  cannot change the probability:

\begin{equation} \label{pbi}
\begin{array}{rl}
p( |a_1 \rangle ;  |\psi_S  \rangle) &= p( |a_1 \rangle ;  |\psi_S  \rangle)_{ \{|a_i \rangle \} }     \\[2.5ex]
&= p( |a_1 \rangle ;  |\psi_S  \rangle)_{ \{|a'_i \rangle \} }    \,.
\end{array}
\end{equation}

\noindent
Now we expressly write the phase of the coefficient $c_1$ in the state of  Eq. (\ref{psit21}):

\begin{equation} \label{psit2ap}
|\psi_{S} \rangle = e^{\phi_1}  |c_{1}| |a_{1} \rangle + |c'_{2}| |a'_{2} \rangle  \,.
\end{equation}

\noindent
We get for the state $|a_1 \rangle$ the probability:

\begin{equation} \label{pg}
p( |a_{1} \rangle ;  |\psi_S  \rangle) = p(  |a_{1} \rangle  ; e^{\phi_1}|c_{1}||a_{1} \rangle + |c'_{2}||a'_{2} \rangle) \,.  
\end{equation}
\noindent
Therefore, considering the dependence $|c_{1}|^2+|c'_{2}|^2=1$ and   that the components of the orthonormal basis  are the constants 
$\langle a'_i| a'_j \rangle=\delta_{ij}$, the most general function of the probability in Eq. (\ref{pg})  can be  written as:  
\begin{equation} \label{pg2}
p( |a_{1} \rangle ;  |\psi_S  \rangle) = p( \phi_1 ,|c_{1}|)  \,.
\end{equation}
\noindent
A similar argument when applied to every state $|a_i \rangle$ gives
\begin{equation} \label{pgi}
p( |a_{i} \rangle ;  |\psi_S  \rangle) = p( \phi_i ,|c_{i}|)  \,.
\end{equation}
\noindent
However, by using the conservation of probability we can show the independence from the phase of $p( |a_i \rangle ;  |\psi_S  \rangle)$. 
Indeed, let us suppose to have two different states $|\psi_{S} \rangle$ and $|\psi_{S}^{\ast} \rangle$ which are distinct just with respect to the  
phase of the component $|a_1 \rangle$
(to simplify the notation, we  select a basis such that the components of the states are different from zero only in a subspace with $dim \mathcal H=2$):
\begin{equation} \label{sphdifbp}
\begin{array}{rl}
|\psi_{S} \rangle &=  e^{\phi_1} |c_{1}| |a_1 \rangle + e^{\phi_2}|c_{2}| |a_2 \rangle           \,,    \\[2.5ex]
|\psi_{S}^{\ast} \rangle &=  e^{\phi_1^\ast} |c_{1}| |a_1 \rangle + e^{\phi_2}|c_{2}| |a_2 \rangle \,,
\end{array}
\end{equation}
\noindent
where with $\phi_1^{\ast}$ we have denoted the new phase.
The conservation of the probability requires:
\begin{equation} \label{cptts0}
\begin{array}{rl}
p (|a_1 \rangle ; |\psi_{S} \rangle) + p(|a_2 \rangle ; |\psi_{S} \rangle  )      &=1          \,,    \\[2.5ex]
p (|a_1 \rangle ; |\psi_{S}^{\ast} \rangle) + p( |a_2 \rangle ; |\psi_{S}^{\ast} \rangle )  &=1   \,,      
\end{array}
\end{equation}
\noindent
where from {\sl Lemma 2} we have used the fact that $p( | a_i \rangle  ;  |\psi_S \rangle)=0$ if  $| a_i \rangle  \bot  |\psi_S \rangle$.
Thanks to Eq. (\ref{pgi}), we have that 
\begin{equation} \label{cptts00}
p(|a_2 \rangle ; |\psi_{S} \rangle  )    = p( |a_2 \rangle ; |\psi_{S}^{\ast} \rangle )  \,.    
\end{equation}
\noindent
From this and Eqs. (\ref {cptts0}),
we also get:
\begin{equation} \label{cptts000}
p(|a_1 \rangle ; |\psi_{S} \rangle)    = p( |a_1 \rangle ; |\psi_{S}^{\ast} \rangle )  \,,
\end{equation}
\noindent
which  is equivalent to write:
\begin{equation} \label{phinf}
p(\phi_1^\ast , |c_{1}|)=p( \phi_1 , |c_{1}|) \,.
\end{equation}
\noindent
Therefore, since $\phi_1^\ast$ is an arbitrary phase, the probability $p(a_1)$ does not depend on it 
but can only depend on the modulus $|c_{1}|$. So,  we can set $p(a_1)=p(|c_{1}|)$, and this proves
our lemma (the generalization to probabilities of other eigenvalues $a_i$ is immediate). 
$\Box$
\vskip.2cm
Now we have all the ingredients to prove our {\sl Theorem 1}: 
the probability $p(a_i)$ is given by the  square modulus of coefficient
$c_i$.
\vskip.2cm
\noindent
{\sl Proof} : 
Let us consider the  state $|\psi_S \rangle \in \mathcal{H}^N$ Eq. (\ref{ps}) with  $N\geq3$. For the sake of simplicity we focus our attention again on $p(a_1)$. 
We choose two bases  $\{|a_i \rangle \}$  and $\{|a'_i \rangle \}$ such that
\begin{subequations} \label{psit2}
\begin{align}
|\psi_{S} \rangle   &=   c_{1}|a_1 \rangle  + c_{2}|a_2\rangle  + c_{3}|a_3\rangle   \label{psit2a} \\[2.5ex]
&=   c_{1}|a_1 \rangle  + c'_{2}|a'_2\rangle       \label{psit2b}               \,,       
\end{align}
\end{subequations}
\noindent
where $c_i=0$ for $i\geq4$, $c'_i=0$ for $i\geq3$ and $|a_1 \rangle =|a'_1 \rangle $.
From the conservation of the total probability  for the states of the two different bases we get respectively:
\begin{subequations} \label{conspt0}
\begin{align}
p(|a_1 \rangle)+p(|a_2\rangle)  +p(|a_3\rangle)    &=1  \,,  \label{conspt0a}   \\[2.5ex]
p(|a_1 \rangle) + p(|a'_2 \rangle)                 &=1  \,,  \label{conspt0b}
\end{align}
\end{subequations}
\noindent
where for NP we have the same probability $p(|a_1 \rangle)$ in Eqs. (\ref{conspt0}).
With a change of variable we can write Eq. (\ref{pci}) as $p(a_i)=f(|c_{i}|^2)$. 
From Eqs. (\ref{conspt0})  it follows that
\begin{equation} \label{conspt}
\begin{array}{rl}
f(|c_{1}|^2)+f(|c_{2}|^2) +f(|c_{3}|^2) &=1   \\[2.5ex]
f(|c_{1}|^2)+f(|c'_{2}|^2) &=1    \,.
\end{array}
\end{equation}
\noindent
By comparing the two Eqs. (\ref{conspt}) we get:
\begin{equation} \label{pla}
f(|c'_{2}|^2)=f(|c_{2}|^2) +f(|c_{3}|^2) \,.
\end{equation}
\noindent
Since the state $|\psi_{S}\rangle$ is normalized, we also have
\begin{subequations} \label{coefn}
\begin{align}
|c_{1}|^2+|c_{2}|^2 +|c_{3}|^2 &=1  \label{coefna} \,,   \\[2.5ex]
|c_{1}|^2+|c'_{2}|^2 &=1            \label{coefnb}  \,,  
\end{align}
\end{subequations}
\noindent
and
\begin{equation} \label{coefnc}
|c'_{2}|^2=|c_{2}|^2 +|c_{3}|^2   \,.
\end{equation}
\noindent
Putting this in Eq. (\ref{pla}):
\begin{equation} \label{pla3}
f(|c_{2}|^2 +|c_{3}|^2)=f(|c_{2}|^2) +f(|c_{3}|^2) \,.
\end{equation}
\noindent
Now we recall that a function $f( x )$ is linear with respect to the variable $x$ if and only if the two following
properties are jointly satisfied:

\begin{equation} \label{lindef}
\begin{array}{rl}
f(x_1+x_2)&=f(x_1)+f(x_2)    \, \quad  \text{additivity} \,,  \\[2.5ex]
f(\alpha x) &= \alpha f(x)    \,\,\,\, \qquad \qquad  \text{homogeneity}   \,,
\end{array}
\end{equation}

\noindent
where $\alpha$ is any real number. 
We remark that if $\alpha$ is a rational number, it can be shown
that the homogeneity follows from the additivity. 
Furthermore, since rational numbers form
a subset dense in the set of  real numbers, such  demonstration can be extended to the case of irrational
$\alpha$, if $f$ is a continuous function \cite{Kuczma}.

Therefore, the condition of additivity is 
sufficient to establish the linearity of the function $f$ if this is continuous. 
On the contrary, if $f$ is not assumed to be  continuous, the condition of additivity
implies linearity only for rational values of the variable $x$.

Hence, if we assume that the probability is continuous with respect to the 
coefficients $|c_{i}|^2$,  Eq. (\ref{pla3}) is a constraint of linearity for the probability.
In this case we can write:
\noindent
\begin{equation}
p(a_i)=k|c_{i}|^2 \,, \quad  i=2,3  \,,
\end{equation}
\noindent
where $k$  denotes a constant.
From  {\sl Lemma 1} we have that  $p(|c_{i}|^2=1)=1$ and therefore $k=1$ 
and $p(a_1)=|c_{1}|^2$.
Hence, the Born rule is deduced for continuous probabilities. 

We point out that  we have postulated the continuity of probability.
However, a demonstration of such a property
from noncontextuality is present in  Gleason's work and it occupies a prominent part of that paper.
Therefore in comparison to Gleason we have only accomplished  half  the job: as Bertrand Russell once said, postulating is equivalent to 
 theft on the honest fatigue. On the other hand, if the continuity of  probability is neither postulated nor demonstrated, 
we can only say that the Born rule applies only for rational values of the coefficients $|c_{i}|^2$.

Thus we now intend to complete our demonstration by showing from NP that the probability
is continuous and that the Born rule also holds  for irrational coefficients.

Consider, once again,  $p(|a_1 \rangle)$. 
We select a basis $\{|a'_i \rangle \}$ such that the state $|\psi_S \rangle$ is written with
two components as in Eq. (\ref{psit2b}),
with  $|a'_1 \rangle \equiv |a_1 \rangle$,
and  the conservation of probability is given by  Eq. (\ref{conspt0b}). 

Then  we choose a new basis $\{|a''_i \rangle \}$  in general with  $|a''_1 \rangle \neq|a_1 \rangle $,
$c''_i =0$ for $ i\geq 4$, but with $|a''_2 \rangle \equiv|a'_2 \rangle $ so that we can write
\begin{equation} \label{ps3}
|\psi_S \rangle= c''_{1}| a''_1 \rangle  + c'_{2}| a'_2 \rangle  +c''_{3}| a''_3 \rangle\,.
\end{equation}
The conservation of probability for the new states of the basis requires that
\begin{equation} \label{cpc2}
p( | a''_1 \rangle)+p( | a'_2 \rangle )+p( | a''_3 \rangle)= 1  \,.
\end{equation}
\noindent
From the previous condition we have the inequality:
\begin{equation} \label{cpc3}
p( | a''_1 \rangle)+p( | a'_2 \rangle) \leq 1  \,.
\end{equation}
\noindent
By comparing Eq. (\ref{conspt0b})  with   Eq. (\ref{cpc3}) and since, thanks to NP,
the probability $p( | a'_2 \rangle)$ is the same in these equations, 
we get
\begin{equation} \label{cpc4}
p( | a''_1 \rangle) \leq p( | a_1 \rangle)   \,.
\end{equation}
\noindent
Finally, we choose a third basis $\{|a'''_i \rangle \}$ such that
$c'''_i =0$ for  $i\geq 4$,   this time with $|a'''_1 \rangle \equiv|a_1 \rangle $:
\begin{equation} \label{ps4}
|\psi_S \rangle= c_{1}| a_1 \rangle  + c'''_{2}| a'''_2 \rangle  +c'''_{3}| a'''_3 \rangle\,.
\end{equation}
\noindent
The conservation of probability gives us:
\begin{equation} \label{cpc5}
p( | a_1 \rangle)+p( | a'''_2 \rangle)+p( | a'''_3 \rangle)= 1  \,.
\end{equation}
\noindent
From Eq. (\ref{cpc5}) we get the inequality:
\begin{equation} \label{cpc6}
p( | a_1 \rangle) \leq 1- p( | a'''_2 \rangle)  \,.
\end{equation}
\noindent
Therefore the probability $p( | a_1 \rangle)$ is in the range:
\begin{equation} \label{cpc7}
p( | a''_1 \rangle) \leq  p( | a_1 \rangle) \leq 1- p( | a'''_2 \rangle)  \,.
\end{equation}
\noindent
Suppose  $|c_{1}|^2$ to be an irrational number, we choose
rational values for $|c''_{1}|^2$ and $|c'''_{2}|^2$ so that
$p( | a''_1 \rangle)=|c''_{1}|^2$
and
$p( | a'''_2 \rangle)=|c'''_{2}|^2$. By  putting  $p( | a_1 \rangle)=p( |c_{1}|^2)$,  Eq (\ref{cpc7}) becomes
\begin{equation} \label{cpc8}
|c''_{1}|^2    \leq  p( |c_{1}|^2 ) \leq 1- |c'''_{2}|^2 \,, \qquad |c''_{1}|^2 \,, |c'''_{2}|^2 \in \mathbb{Q} \,,
\end{equation}
\noindent
with $|c''_{1}|^2 \leq 1- |c'''_{2}|^2$. 
Let us denote with $|\psi_{S\bot} \rangle$ the  orthonormal state to $|\psi_{S} \rangle$.
Consider the limits  $| a''_1 \rangle \rightarrow | a_1 \rangle $, $| a''_3 \rangle \rightarrow |\psi_{S\bot} \rangle $    and  
$| a'''_2 \rangle \rightarrow | a'_2 \rangle $, $| a'''_3 \rangle \rightarrow |\psi_{S\bot} \rangle$. 
We have $|c''_{1}|^2 \rightarrow |c_{1}|^2$, $|c''_{3}|^2 \rightarrow 0$ and  $|c'''_{2}|^2 \rightarrow |c'_{2}|^2$, $|c'''_{3}|^2 \rightarrow 0 $. 
Summarizing, with the normalization condition  in Eq. (\ref{coefnb}) we get:
\begin{equation} \label{lim9}
|c''_{1}|^2 \rightarrow |c_{1}|^2 \,, \quad 1- |c'''_{2}|^2 \rightarrow |c_{1}|^2 \,.
\end{equation}
\noindent
Because  rational numbers form a dense set in the set of real numbers, we can always find a couple of numbers
$|c''_{1}|^2 \,, 1-|c'''_{2}|^2 \in \mathbb{Q}$
which are as near as we want to the irrational number $|c_{1}|^2$. Therefore, from Eq. (\ref{cpc8}) it follows that
the oscillation of the function $p$ 
is zero at $|c_{1}|^2$. Consequently,  the probability must be continuous at $|c_{1}|^2$ and $p( |c_{1}|^2 ) =|c_{1}|^2$ even if $|c_{i}|^2 \notin \mathbb{Q}$.
$\Box$

\section{Noncontextual probability for entangled states}

 {\sl Theorem 1} with NP  can be generalized  to include also
entangled states, bipartite or multipartite systems and quantum measurement devices.

Let us indicate with $\mathcal C$ the context, including
our experimental devices and the environment, with
${|C_i \rangle} \in \mathcal{H}^M$ its possible states.
Consider again the measurement of the two observables  $\hat{A}$ and $\hat{A}'$ introduced in the proof of {\sl Lemma 3} in the
previous section.
The choice of measuring either the eigenvalues $\left\{a_2, a_3 \right\}$ or $\left\{a'_2, a'_3 \right\}$  together with $a_1$  
corresponds to two different experimental arrangements $\textsf{A}$ and $\textsf{A'}$.

Suppose that at time $t_0$ the system is in the initial state 
$|\psi_{SC}(t_0) \rangle = |\psi_{S} \rangle |C_0 \rangle$.
At time $t$, after the interaction with the measurement device, the entanglement between the system $\mathcal S$ with the experimental setup  
$\mathcal C$  correlates the observable $\hat{A}$ (or $\hat{A}'$) of the system with the pointers  of
the experimental device. If we measure $\hat{A}$ we have the state:
\begin{equation}   \label{psxinena}
|\psi_{SC}(t) \rangle = c_{1}| a_1 \rangle | C_{1}\rangle+c_{2}| a_2 \rangle | C_{2} \rangle+c_{3}| a_3 \rangle | C_{3} \rangle \,,
\end{equation}
\noindent
while if we measure $A'$ we get:
\begin{equation} \label{psxinenaa}
|\psi_{SC'}(t) \rangle = c_{1}| a_1 \rangle | C'_{1}\rangle
 +c'_{2}| a'_2 \rangle | C'_{2} \rangle
+c'_{3}| a'_3 \rangle | C'_{3} \rangle \,.
\end{equation}
The global wave functions Eqs. (\ref{psxinena}-\ref{psxinenaa}) 
are different in the different contexts $\textsf{A}$ and $\textsf{A'}$. So 
the two probabilities
\begin{equation} \label{cpgs}
\begin{array}{rl}
p_{A}(a_1)  &= p(|a_1 \rangle | C_{1}\rangle ; | \psi_{SC}(t) \rangle)  \,,  \\[2.5ex]       
p_{A'}(a_1) &= p(|a_1 \rangle | C'_{1}\rangle ; | \psi_{SC'}(t) \rangle)  \,,
\end{array}
\end{equation}
may be different.  
The possible contextuality of probability is now included inside the description of the global system ${\mathcal S} + {\mathcal C}$
of the quantum state.
We notice that it is no longer possible to write
\begin{equation} \label{hd}
p_{A}(a_1)  = p(|a_1 \rangle | C_{1}\rangle ; | \psi_{SC}(t) \rangle , \{ | a_1 \rangle, | a_2 \rangle, | a_3 \rangle \})  \,,
\end{equation}
if we accept the completeness of QM. Otherwise, indeed, 
some  parameters would be outside the description of the  state $| \psi_{SC}(t) \rangle$, which now englobes not only 
the system but also the context of the measurement. 
This should  imply the existence of  hidden variables.

Assuming NP requires the equality between the probabilities in  Eq. (\ref{cpgs}):
\begin{equation} \label{ncpent}
p(|a_1 \rangle | C_{1}\rangle ; | \psi_{SC}(t) \rangle)\equiv p(|a_1 \rangle | C'_{1}\rangle ; | \psi_{SC'}(t) \rangle)  \,,
\end{equation}

\noindent
which generalizes the condition in Eq. (\ref{ncps}). In fact, Eq. (\ref{ncpent}) by the property of perfect correlation becomes equivalent to Eq. (\ref{ncps}).

In a similar way as in the previous demonstration, 
by using the state of Eq. (\ref{psxinena}) and a state $|\psi_{S} \rangle$ 
with a new basis $ \{ | a'_1 \rangle \}$   such that $c'_i=0$ for $i\geq 3$, $| a'_1 \rangle \equiv| a_1 \rangle$ we can write:
\begin{equation} \label{psce2}
|\psi_{SC'}(t) \rangle = c_{1}| a_1 \rangle | C'_{1}\rangle
 +|c'_{2}|| a'_2 \rangle | C'_{2} \rangle  \,.
\end{equation}
From the condition of NP it is not difficult to deduce that
$p(|a_i \rangle | C_{i}\rangle ; | \psi_{SC}(t) \rangle) =p(|c_{i}|)$ must hold. Then, employing  states like
in Eq. (\ref{psxinena}) and Eq. (\ref{psce2}) and from NP, normalization condition and conservation of probability, it is possible to prove again the linearity of the probabilities with respect to the square moduli of the amplitudes (continuity can also be generalized to the present case).
Putting all pieces together, we get
\begin{equation} \label{bren2}
\begin{array}{rl}
p(|a_i \rangle | C_{i}\rangle ; | \psi_{SC}(t) \rangle) &= |c_i|^2      \\[2.5ex]       
&= |\langle a_i | \langle  C_{i} |\, | \psi_{SC}(t) \rangle  |^2       \\[2.5ex]
&=|\langle a_i |  \psi_{S} \rangle  |^2  \,.
\end{array}
\end{equation}
We note that the state of the global system $\mathcal{S}+\mathcal{C}$ is in the Hilbert space  $\mathcal{H}^N \otimes\mathcal{H}^M$ with dimension
$N \cdot M$. If $M\geq3$ this allows us to include in the demonstration of the Born rule also states with $N=2$. In fact, 
consider a system $\mathcal{S}$ described by a state $|\psi_{S}  \rangle$ in a Hilbert space with $dim \mathcal H=2$ and 
basis $\{|a_1 \rangle, |a_2 \rangle \}$. With a measurement we could use the context $\mathcal C$  as an ancilla system to build  the following entangled states:
\begin{equation} \label{psith2}
\begin{array} {rl}
|\psi_{SC}^0  \rangle   &=   c_{1}|a_1 \rangle |C_1 \rangle + c_{2}|a_2 \rangle |C_2 \rangle  \,, \\[2.5ex]
|\psi_{SC'}^1  \rangle   &=   c_{1}|a_1 \rangle |C'_1 \rangle + c'_{2}|a_2 \rangle |C'_2 \rangle
+ c'_{3}|a_2 \rangle |C'_3 \rangle \,,
\end{array}
\end{equation}
where $|\psi_{SC'}^1  \rangle $ is obtained by dividing the part of the state $|\psi_{S}  \rangle$ associated  with the eigenvalue $a_2$ into two regions. 
In a similar way it is possible to demonstrate that $p(a_i)=|\langle a_i |  \psi_{S} \rangle  |^2$ as in the previous cases.   
\section{Noncontextual probability and resolution of the measurement}
A simple application of  {\sl Lemma 3} shows
that equal amplitudes are related to equal probabilities, i.e.: if $c_i=c_j$ with
$i\neq j$ then $p(a_i)=p(a_j)$.
Whereas {\sl Lemma 3} is derived from the NP principle, the same 
result was obtained by Zurek and Deutsch from two different invariance principles. 
From here both authors study the more general case 
with states having different rational amplitudes using the fine-graining technique.
We shortly summarize their results. We consider the fine-graining in Zurek's approach with the environment in a  simple situation (our remarks  also hold for Deutsch's similar derivation). We  assume that the state of the system ${\mathcal S}$ is in a Hilbert space with $dim{\mathcal H}=2$, entangled with the the state of environment ${\mathcal E}$ such that their joint state is 
\begin{equation} \label{fine1}
|\psi_{\mathcal{SE}} \rangle=   \frac{1}{\sqrt{N}} |a_{1}\rangle  |\varepsilon_{1}\rangle  + 
\sqrt{\frac{N-1}{N}}  |a_{2}\rangle  |\varepsilon_{2}\rangle        \,.
\end{equation}
The trick is to transform this state with unequal coefficients into a state with equal coefficients by  fine-graining.
This goal is reached by extracting from the  environment an ancilla system $\mathcal C$ having states   
\begin{equation} \label{coanc}
|C'_{1} \rangle=    |C_{1}\rangle \,,  \qquad  |C_{2} \rangle = \sum^N_{i=2} |C'_{i}\rangle /\sqrt{N-1}        \,,
\end{equation}
which become correlated with the states $\{|a_{1}\rangle,|a_{2}\rangle \}$ of $\mathcal S$. Within a quantum measurement, $\mathcal C$ can be regarded as a counter and $\{|C'_{i}\rangle \}$ as a new orthonormal basis of  pointer states. Denoting with $\{|e_{i}\rangle \}$ the new states of the environment without $\mathcal C$, after the interaction between $\mathcal S$  and $\mathcal C$ we have the joint system
\begin{equation} \label{entanc}
|\psi_{\mathcal{SCE}} \rangle = \frac{1}{\sqrt{N}} |a_{1}\rangle |C_{1}\rangle  |e_{1}\rangle  + 
 \sqrt{\frac{N-1}{N}}|a_{2}\rangle |C_{2}\rangle  |e_{2}\rangle   \,,          
\end{equation}
which can be expressed with the new basis of pointer states of Eq. (\ref{coanc}):
\begin{equation} \label{entanc2}
|\psi_{\mathcal{SC'E}} \rangle = \frac{1}{\sqrt{N}} |a_{1}\rangle |C_{1}\rangle  |e_{1}\rangle  
+ \frac{1}{\sqrt{N}} \sum^N_{i=2} |a_{2}\rangle |C'_{i}\rangle  |e_{2}\rangle    \,.
\end{equation}
Now we have a new state with equal amplitudes. 
But how can we be sure that after  
fine-graining  the probability $p(|a_{1}\rangle)$ with the state $|\psi_{\mathcal{SC'E}} \rangle$ is the same with the state $|\psi_{\mathcal{SCE}} \rangle$? In principle they could be different, so to legitimately use
the fine-graining reasoning  we have to assume the equality:
\begin{equation} \label{finncp}
p(|a_{1}\rangle ; |\psi_{\mathcal{SC'E}} \rangle )=p(|a_{1}\rangle ;|\psi_{\mathcal{SCE}} \rangle )       \,,
\end{equation}
equivalent to the condition of NP encountered in Eq. (\ref{ncpent}) of section 3.  Then, in their demonstrations Zurek and Deutsch implicitly assume NP.
We notice that fine-graining is equivalent to a change of measurement resolution, 
and this has been used in our derivation as well. 
A change of resolution, for instance, 
is considered going  from Eq. (\ref{conspt0a}) to Eq. (\ref{conspt0b}).  There we assumed from NP 
that $p(|a_{1}\rangle)$  is the same in the two equations and therefore: 
\begin{equation} \label{invpris}
p(|a'_{2}\rangle) =  p(|a_{2}\rangle) +p(|a_{3}\rangle)    \,,
\end{equation}
where $p(|a'_{2}\rangle)$ is the probability of finding the state of the system in the subspace orthonormal to $|a_{1}\rangle$, 
whereas $p(|a_{2}\rangle)$ and $p(|a_{3}\rangle)$ are the probabilities of finding it, respectively, in the subspaces 
$|a_{2}\rangle$ and $|a_{3}\rangle$ of the subspace orthonormal to $|a_{1}\rangle$. In fact, in terms of projectors, Eq. (\ref{invpris}) can be written as Gleason's Postulate 2:
\begin{equation} \label{invprispr}
p(|a_{2}\rangle \langle a_{2} | + |a_{3}\rangle \langle a_{3} |) =  
p(|a_{2}\rangle \langle a_{2} |) + p(|a_{3}\rangle \langle a_{3}|)   \,.
\end{equation}
Thus NP could be physically interpreted  as a probability invariance when we change the resolution of the measurement connected with other properties of the system. This interpretation gives physical insight to the NP principle as an invariance for the resolution change.

\section{Noncontextual probability and no-faster-than-light signalling}
In general, every measurement in the real world happens in the spacetime. Every measurement of any observable is necessarily  reduced   to a measurement of position at a certain time. 
For instance, the Stern-Gerlach
apparatus correlates a spin observable with a position observable so that different values of the spin univocally correspond to different position values of the system. A measurement of position detects the spin value of the system.

Let us introduce in our representation of $\mathcal S$ the observable position $\hat{X}$.
We assume that at time $t_0$, the system is in the initial state $|\psi_{SC}(t_0) \rangle = |\psi_{S} \rangle |x_0 \rangle |C_0 \rangle$.
For simplicity of notation,  from now on we denote with $|C_{xi} \rangle=|x_i \rangle|C_i \rangle$ the  position state of the system 
with its context, where the  letter $x$  will remind us the dependence on the position of the system and the experimental apparatus. 

We measure the observable $\hat{A}$ or $\hat{A}'$. If we measure $\hat{A}$ we get:
\begin{equation}   \label{psxinenab}
|\psi_{SC}(t) \rangle = c_{1}| a_1 \rangle | C_{x1}\rangle
+c_{2}| a_2 \rangle | C_{x2} \rangle
+c_{3}| a_3 \rangle | C_{x3} \rangle \,,
\end{equation}
while if we measure $\hat{A}'$ we have the different state:
\begin{equation} \label{psxinenaab}
|\psi_{SC'}(t) \rangle = c_{1}| a_1 \rangle | C'_{x1}\rangle
 +c'_{2}| a'_2 \rangle | C'_{x2} \rangle
+c'_{3}| a'_3 \rangle | C'_{x3} \rangle \,.
\end{equation}
\noindent
We want to prove the following
\vskip.2cm
\noindent
{\sl Theorem 2} : If we assume the no-faster-than-light signalling condition for measurement
events which are spacelike separate, then the probability $p(a_i)$ must be  noncontextual.
\vskip.2cm

\noindent
{\sl Proof}: If the measurements of $\left\{a_2, a_3 \right\}$ or $\left\{a'_2, a'_3 \right\}$ at the points $x_2$ and $x_3$ are events spacelike separate with respect to the measurement of $a_1$ at $x_1$, according to the relativistic hypothesis of no-signalling, the probability $p(a_1)$
of getting the eigenvalue $a_1$ has to be the same in both  different experimental contexts \textsf{A} and \textsf{A'}. Otherwise an observer situated in the place of measurement of  $\left\{a_2, a_3 \right\}$ and $\left\{a'_2, a'_3 \right\}$ would be able to send a superluminal signal to another observer situated in the place where $a_1$ is measured. Then, according to the no-faster-than-light signalling hypothesis:
\begin{equation} \label{stncp2b}
p(|a_1 \rangle | C_{x1}\rangle ; | \psi_{SC}(t) \rangle)
= p(|a_1  \rangle | C'_{x1}\rangle ; |\psi_{SC'}(t) \rangle) \,,
\end{equation}
which is to say, with the result of Eq. (\ref{pacbsd}):
\begin{equation} \label{ncp2b}
\begin{array}{rl}
p(a_1) &= p(|a_1 \rangle ;|\psi_{SC}(t) \rangle)   \\[2.5ex]
&= p(|a_1 \rangle ;|\psi_{SC'}(t) \rangle) \,.
\end{array}
\end{equation}
Therefore  $p(a_1)$ must be noncontextual. $\Box$

Consider, beside the measurement of the eigenvalues $\{ a_1, a_2, a_3 \}$ of the observable $\hat{A}$, also the simple possibility of measuring whether our system has the eigenvalue $a_1$ or not, which
is equivalent to check  whether the system  $\mathcal S$ is  in the state $|a_1 \rangle$ or in the subspace $|a_{1\bot} \rangle$ orthogonal to it.
In such a case, during the process of measurement, the system $\mathcal S+C$  will be in the state: 
\begin{equation} \label{contglns}
|\psi_{SX_{\bot}}(t) \rangle = c_{1}| a_1 \rangle | C_{x1}\rangle
 +c_{2\bot}| a_{1\bot} \rangle | C_{x2\bot} \rangle  \,.
\end{equation}
If the two choices of measuring either  
$\{ |a_1\rangle, |a_2\rangle, |a_3 \rangle \}$ or $\{ |a_1\rangle, |a_{1\bot}\rangle \}$ 
correspond to events spacelike separate, as mentioned above, from the condition of no-signalling we must have the NP condition: 
\begin{equation} \label{ncgl}
\begin{array}{rl}
p(a_1) &= p(|a_1 \rangle ;|\psi_{SC}(t) \rangle)   \\[2.5ex]
&= p(|a_1 \rangle ; |\psi_{SC_{\bot}}(t) \rangle ) \,.
\end{array}
\end{equation}
From Eq. (\ref{ncgl}) the conservation of  probability imposes also the equation: 
\begin{equation} \label{ncglcg}
\begin{array}{rl}
p(|a_{1\bot}\rangle ; &|\psi_{SC_{\bot}}(t) \rangle )     =   \\[2.5ex]
&= p(|a_2 \rangle ;|\psi_{SC}(t) \rangle) +p(|a_3 \rangle ;|\psi_{SC}(t) \rangle) \,.
\end{array}
\end{equation}
How  we showed at the end of Section IV, Eq. (\ref{ncglcg}) is equivalent to the initial hypothesis adopted by Gleason.
Moreover, by examining Eq. (\ref{ncgl}) we understand that it is analogous
to assuming an invariance condition of the probability under a resolution change of our measurement, or fine/coarse graining.
So, at least for spacelike separate measurements, NP could be deduced by a relativistic principle.
However, the Born rule holds for general events, not only for  spacelike separate ones. Hence NP or, equivalently, the invariance of probability when we change resolution, seem to be  more general principles. 
\section{Concluding remarks}
We have given a new demonstration of the Born rule based on noncontextual probability and on some non-statistical postulates of QM. The introduction of a pure quantum state of the system has allowed
a more elementary proof than Gleason's theorem. This may seem, at first sight, a loss of generality 
with respect to Gleason's proof, however Gleason  also introduces vectors of basis which can be regarded as  states of the system. Moreover, the step from pure states to density operators and the trace rule for a mixed ensemble is quite natural when we have the Born rule in hand \cite{Sakurai}.

As in Gleason's, our demonstration holds for   $dim \mathcal H \geq3$, where   $dim \mathcal H$ could be finite or infinite. The origin of this dimensional limitation lies in the fact that for $dim \mathcal H \geq3$ the same state vector may belong to different  bases. This, with NP and the normalization of the probability and  the quantum state,  brings us to a constraint of functional linearity between the probability and the amplitude square modulus.
Our theorem can be simply generalized in order to include multipartite states. In such a case, if the state is $dim \mathcal H \geq6$, the Born rule can be demonstrated to hold also in subspaces with  $dim \mathcal H=2$.

We have also given a demonstration that the probability is continuous on the amplitudes using NP. 
 
Physical motivation for  NP could come from  a relativistic principle of no-faster-than-light signalling. However, the Born rule also holds for measurements which are non-spacelike separate, hence NP seems to be
a more fundamental principle. Finally, we have given an interpretation  of NP as an 
invariance of the probability under a resolution change of our measurement.

\appendix
\section{}
\label{lemma1}

\noindent
We want to generalize the result of {\sl Lemma 1} for entangled states.
Suppose that the system is prepared with  eigenvalues $\left\{a_i, b_k\right\}$
and therefore with eigenstates $\left\{|a_i \rangle , |b_k \rangle \right\}$. Similarly to the
unentangled case, we can deduce that
$p(\left\{a_j, b_l \right\}) =\delta_{ij}\delta_{kl}$ if and only if
$p( \left\{|a_j \rangle , |b_l \rangle \right\}) = \delta_{ij} \delta_{kl}$. This relation,
 translated into conditional probabilities,  is equivalent to write:
\begin{equation} \label{picond2d1}
\begin{array}{rl}
p(\left\{a_i, b_k \right\} ; \{|a_j \rangle , |b_l \rangle \}) &= \\[2.5ex]
p(\{|a_j \rangle , |b_l \rangle \}; \left\{a_i, b_k \right\}) &=\delta_{ij} \delta_{kl} \,.
\end{array}
\end{equation}
Then, from Eq. (\ref{picond2d1}) we have that for a generic state as in  Eq. (\ref{psc}):
\begin{equation} \label{bfmac1}
\begin{array}{rl}
 p(\{a_i, b_k \}) &= {\displaystyle \sum_{j=1}^N} {\displaystyle \sum_{l=1}^M}\,
p( \{a_i, b_k \} , \{ |a_j \rangle  , | b_l \rangle \} ) \\[2.5ex]
= {\displaystyle \sum_{j=1}^N} {\displaystyle \sum_{l=1}^M}\,&
p( \{a_i, b_k \} \,; \{ |a_j \rangle ,| b_l \rangle \})  \, p( \{| a_j \rangle ,  | b_l \rangle \}) \\[2.5ex]
&=p( \{| a_i \rangle ,  | b_k \rangle \})   \,.
\end{array}
\end{equation}
If the observable $\hat{A}$ has an eigenstate $|a_i \rangle$ and the observable $B$ has  
an eigenstate $|b_k \rangle$ then the tensorial product  $|a_i \rangle |b_k \rangle$ is an eigenstate
of the observable $\hat{A}\otimes \hat{B}$, and vice versa. The eigenstates $\left\{|a_i \rangle |b_k \rangle\right\}$ of $\hat{A}\otimes \hat{B}$
form a basis of Hilbert space  $\mathcal{H}^N \otimes \mathcal{H}^M$. If we indicate with
$p(|a_j \rangle |b_l \rangle)$ the probability of obtaining the eigenstate  $|a_j \rangle |b_l \rangle$, with a measurement of $\hat{A}\otimes \hat{B}$, we have $p( \left\{|a_j \rangle , |b_l \rangle \right\}) = \delta_{ij}\delta_{kl}$   if and only if
$p( |a_j \rangle |b_l \rangle) = \delta_{ij} \delta_{kl}$, or, in equivalence:
\begin{equation} \label{picond2d}
\begin{array}{rl}
p(\{ | a_i \rangle , | b_k \rangle  \} \,; |a_j \rangle  |b_l \rangle) &= \\[2.5ex]
p(|a_j \rangle  |b_l \rangle ; \{|a_i\rangle  , |b_k \rangle  \}) &=\delta_{ij} \delta_{kl} \,.
\end{array}
\end{equation}
From Eqs. (\ref{bfmac1})-(\ref{picond2d}), through the use of the conditional probabilities, it immediately
follows that on the  state $|\psi_{SB} \rangle$ we have $p( | a_i \rangle  | b_k \rangle )
=  p(\left\{a_i, b_k \right\})$.
\vskip.2cm

\end{document}